\documentclass[12pt]{article}
\usepackage{graphicx}
\usepackage{amsmath}
\usepackage{amssymb}
\usepackage{caption2}
\begin{document}
\bibliographystyle{prsty}
\begin{center}
{\large {\bf \sc{ Analysis the $0^{++}$ nonet mesons
as four-quark states with  the QCD sum rules }}} \\[2mm]
Zhi-Gang Wang$^{1}$ \footnote{Corresponding author;
E-mail,wangzgyiti@yahoo.com.cn. },
Wei-Min Yang$^{2}$ and Shao-Long Wan$^{2}$  \\
$^{1}$ Department of Physics, North China Electric Power University, Baoding 071003, P. R. China \\
$^{2}$ Department of Modern Physics, University of Science and Technology of China, Hefei 230026, P. R. China \\
\end{center}

\begin{abstract}
In this article, we take the point of view  that the $0^{++}$
nonet mesons below $1 GeV$ are diquark-antidiquark states
$(qq)_{\bar{3}}(\bar{q}\bar{q})_3$, and devote to determine their
masses in the framework of the QCD sum rules approach with the
interpolating currents constructed from scalar-scalar type and
pseudoscalar-pseudoscalar type diquark pairs respectively. The
numerical results indicate that the $0^{++}$ nonet mesons  may
have two possible diquark-antidiquark substructures.
\end{abstract}

PACS numbers:  12.38.Lg; 13.25.Jx; 14.40.Cs

{\bf{Key Words:}}  Scalar nonet mesons,  QCD sum rules
\section{Introduction}
The light flavor  scalar mesons present a remarkable exception for
the constituent  quark model and the structures of those mesons
have not been unambiguously determined yet \cite{Godfray}.
Experimentally, the strong overlaps with each other
 and the broad widths ( for the $f_0(980)$, $a_0(980)$ et al, the widths are comparatively narrow)
 make their spectra cannot be approximated by the
  Breit-Wigner  formula.  The numerous
candidates  with the same quantum numbers $J^{PC}=0^{++}$ below $2
GeV$ can not be accommodated in one $q\bar{q}$ nonet,  some are
supposed to be glueballs, molecules and multiquark states
\cite{JaffeAchasov4q,IsgurKK,HDress,Close2002,ReviewScalar}. There
maybe different dynamics dominating the $0^{++}$ mesons  below and
above $1GeV$ which results in  two scalar nonets below $1.7 GeV$.
The attractive interactions of one-gluon exchange favor the
formation of diquarks in  color antitriplet $\overline{3}_{ c}$,
flavor antitriplet $\overline{3}_{ f}$ and spin singlet  $ 1_{s}
$.  The strong attractions between the states
$(qq)_{\overline{3}}$ and $(\bar{q}\bar{q})_{3}$  in $S$-wave may
result in a nonet manifested below $1GeV$ while the conventional
$^3P_0$ $\bar{q}q$ nonet would have masses about $1.2-1.6 GeV$. In
the same energy region, there are two well established siblings
$^3P_1$ and $^3P_2$ $\bar{q}q$ nonets with $J^{PC}=1^{++}$ and
$2^{++}$ respectively. Furthermore, there are enough candidates
for the $^3P_0$ $\bar{q}q$ nonet mesons , $a_0(1450)$,
$f_0(1370)$, $K^*(1430)$, $f_0(1500)$ and $f_0(1710)$
\cite{Close2002,ReviewScalar}.

Taking  the diquarks and antidiquarks as the basic constituents,
keeping  the effects of the $s$ quark mass at the first order, the
two isoscalars $\bar u\bar d u d$ and $\bar s s\frac{\bar u u+\bar
d d}{\sqrt{2}}$ mix ideally, the $\bar s s\frac{\bar u u+\bar d
d}{\sqrt{2}}$
 degenerate with the isovectors  $\bar s s\bar d u$,
$\bar s s\frac{\bar u u-\bar d d}{\sqrt{2}}$ and $\bar s s\bar u
d$ naturally. Comparing with  the traditional ${\bar q} q $ nonet
mesons, the mass spectrum is inverted.  The lightest state is the
non-strange isosinglet ($\bar u\bar d ud$), the heaviest are the
degenerate isosinglet and isovectors with hidden $\bar s s$ pairs
while the four strange states lie in between
\cite{Close2002,ReviewScalar}.
  The broad isosinglet  $S$-wave $\pi\pi$ resonance near $600MeV$ can be signed
  to be the $f_{0}(600)$ or $\sigma$ meson with quark constituent
   $\bar u\bar d ud$. The well established isoscalar $f_{0}(980)$ and isovector
$a_{0}(980)$ mesons which lie just below the $\overline KK$
threshold have quark constituents  $\bar s s\frac{\bar u u+\bar d
d}{\sqrt{2}}$ and $\bar s s\frac{\bar u u-\bar d d}{\sqrt{2}}$
respectively in the four-quark model. The four light
isospin-$\frac{1}{2}$ $K\pi$ resonances near $800 MeV$, known as
the $\kappa(800)$  mesons, have not been conformed yet, there are
still controversy about their existence \cite{kappa}.

In this article, we take the point of view that the scalar
$0^{++}$ nonet mesons below $1GeV$ are  four-quark states
$(qq)_{\bar{3}}(\bar{q}\bar{q})_3$ in the ideal mixing limit, and
devote to determine the values of their masses  in the framework
of  the  QCD sum rules approach
\cite{Shifman79,Maiani04,Nielsen04,WangY05}.

The article is arranged as follows: in section II, we obtain the
QCD sum rules for the masses of the $0^{++}$ nonet mesons; in
section III, numerical results; section IV is reserved for
conclusion.

\section{ Masses of the $0^{++}$ nonet mesons  with the  QCD Sum Rules}

In the four-quark models, the structures of the scalar nonet
mesons in the ideal mixing limit can be symbolically taken as
\cite{JaffeAchasov4q,Close2002,ReviewScalar}
\begin{eqnarray}
\sigma(600)=ud\bar{u}\bar{d},\;&&f_0(980)={us\bar{u}\bar{s}+
ds\bar{d}\bar{s}\over\sqrt{2}}, \nonumber\\
a_0^-(980)=ds\bar{u}\bar{s},\;&&a_0^0(980)={us\bar{u}\bar{s}
-ds\bar{d}\bar{s}\over\sqrt{2}},\;a_0^+(980)=us\bar{d}\bar{s},
\nonumber\\
\kappa^+(800)=ud\bar{d}\bar{s},\;&&\kappa^0(800)=ud\bar{u}\bar{s},\;\;\;\;\;\;
\bar{\kappa}^0(800)=us\bar{u}\bar{d},
\;\;\;\;\;\;\kappa^-(800)=ds\bar{u}\bar{d}. \nonumber
\end{eqnarray}
The four-quark configurations of the  $J^{PC}=0^{++}$ mesons can
give a lot of satisfactory  descriptions of the hadron phenomenon,
for example, the mass degeneracy of the $f_0(980)$ and $a_0(980)$
mesons, the mass hierarchy pattern of the scalar nonet, the large
radiative widths of the $f_0(980)$ and $a_0^0(980)$ mesons, the
$D^+$ to $\pi^+\pi^+\pi^-$ and $D_s^+(c\bar{s})$ to
$\pi^+\pi^+\pi^-$ decays.

In the following, we write down the interpolating currents  for
the scalar nonet mesons below $1GeV$ based on the four-quark model
\cite{Maiani04,Nielsen04,WangY05},
\begin{eqnarray}
J^A_{f_0}&=&\frac{\epsilon^{abc}\epsilon^{ade}}{\sqrt{2}}\left[(u_b^TC\gamma_5s_c)
(\bar{u}_d\gamma_5C\bar{s}_e^T)+(d_b^TC\gamma_5s_c)
(\bar{d}_d\gamma_5C\bar{s}_e^T)\right], \\
J^B_{f_0}&=&\frac{\epsilon^{abc}\epsilon^{ade}}{\sqrt{2}}\left[(u_b^TCs_c)
(\bar{u}_dC\bar{s}_e^T)+(d_b^TC s_c) (\bar{d}_d
C\bar{s}_e^T)\right], \\
J^A_{a_0^0}&=&\frac{\epsilon^{abc}\epsilon^{ade}}{\sqrt{2}}\left[(u_b^TC\gamma_5s_c)
(\bar{u}_d\gamma_5C\bar{s}_e^T)-(d_b^TC\gamma_5s_c)
(\bar{d}_d\gamma_5C\bar{s}_e^T)\right], \\
J^B_{a_0^0}&=&\frac{\epsilon^{abc}\epsilon^{ade}}{\sqrt{2}}\left[(u_b^TCs_c)
(\bar{u}_dC\bar{s}_e^T)-(d_b^TCs_c)
(\bar{d}_dC\bar{s}_e^T)\right],\\
J^A_{a_0^+}&=&\epsilon^{abc}\epsilon^{ade}(u_b^TC\gamma_5s_c)
(\bar{d}_d\gamma_5C\bar{s}_e^T), \\
J^B_{a_0^+}&=&\epsilon^{abc}\epsilon^{ade}(u_b^TCs_c)
(\bar{d}_dC\bar{s}_e^T), \\
J^A_{\kappa^+}&=&\epsilon^{abc}\epsilon^{ade}(u_b^TC\gamma_5d_c)
(\bar{d}_d\gamma_5C\bar{s}_e^T), \\
J^B_{\kappa^+}&=&\epsilon^{abc}\epsilon^{ade}(u_b^TCd_c)
(\bar{d}_dC\bar{s}_e^T), \\
J^A_{\kappa^0}&=&\epsilon^{abc}\epsilon^{ade}(u_b^TC\gamma_5d_c)
(\bar{u}_d\gamma_5C\bar{s}_e^T), \\
J^B_{\kappa^0}&=&\epsilon^{abc}\epsilon^{ade}(u_b^TCd_c)
(\bar{u}_dC\bar{s}_e^T), \\
J^A_{\sigma}&=&\epsilon^{abc}\epsilon^{ade}(u_b^TC\gamma_5d_c)
(\bar{u}_d\gamma_5C\bar{d}_e^T), \\
J^B_{\sigma}&=&\epsilon^{abc}\epsilon^{ade}(u_b^TCd_c)
(\bar{u}_dC\bar{d}_e^T),
\end{eqnarray}
where $a,~b,~c,~...$ are color indices and $C$ is the charge
conjugation matrix.  The constituents $S^a(x) = \epsilon^{abc}
u_b^T(x)C\gamma_5 d_c(x),\epsilon^{abc} u_b^T(x)C\gamma_5
s_c(x),\epsilon^{abc} d_b^T(x)C\gamma_5 s_c(x) $ and $P^a(x) =
\epsilon^{abc} u_b^T(x)Cd_c(x),\epsilon^{abc}
u_b^T(x)Cs_c(x),\epsilon^{abc} d_b^T(x)Cs_c(x)  $ represent the
scalar $J^P=0^+$ and the pseudoscalar $J^P=0^-$  diquarks
respectively. They both belong to the antitriplet $\bar{3}$
representation of the color $SU(3)$ group and can cluster together
to form $S^a-\bar{S}^a$ type and $P^a-\bar{P}^a$ type diquarks
pairs to give the correct spin and parity for the scalar mesons
$J^P=0^+$ . The scalar diquarks correspond to the $^1S_0$ states
of $ud$, $us$ and $ds$ diquark systems. The one-gluon exchange
force and the instanton induced force can lead to significant
attractions between the quarks in the $0^+$ channels
\cite{GluonInstanton}. The pseudoscalar diquarks do not have
nonrelativistic limit,  can be  taken as  the $^3P_0$ states. As
the instanton induced force results in strong attractions in the
scalar diquark channel and strong repulsions in the pseudoscalar
diquark channel, if the effects of the instanton are manifested in
the $0^{++}$ nonet mesons, we shall prefer the $S^a-\bar{S}^a$
type interpolating currents to the $P^a-\bar{P}^a$ type
interpolating currents \cite{GluonInstanton,Wang05}.

  The calculation of the  $a_0(980)$ meson as a four-quark
state in the QCD sum rules approach  was done  originally for the
 decay constant and the hadronic coupling constants
with the interpolating currents $J_{a_0}^1$ and $J_{a_0}^2$
\cite{Latorre85,Narison86}\footnote{There is also other work based
on the four-quark model with QCD sum rules \cite{Braun89},
however, it is not available in NCEPU. },
\begin{eqnarray}
J_{f_0(a_0)}^1&=&\Sigma_{\Gamma=1,\pm \gamma_5} \bar{s}\Gamma
s\frac{\bar{u}\Gamma u\pm \bar{d}\Gamma d}{\sqrt{2}}\, , \nonumber \\
J_{f_0(a_0)}^2&=&\Sigma_{\Gamma=1,\pm \gamma_5} \bar{s}\Gamma
\frac{\lambda^a}{2} s\frac{\bar{u}\Gamma \frac{\lambda^a}{2} u\pm
\bar{d}\Gamma \frac{\lambda^a}{2} d}{\sqrt{2}},
\end{eqnarray}
where the $\lambda^a$ is the $SU(3)$ Gell-Mann matrix.   Perform
Fierz transformation both in the Dirac spinor and color space, for
example, we can obtain
 \begin{eqnarray}
J_{f_0}^2 &\propto& C_A
J^A_{f_0}+C_BJ^B_{f_0}\cdots \, , \nonumber \\
J_{a_0}^2 &\propto& C_A J^A_{a_0}+C_BJ^B_{a_0}\cdots \, .
 \end{eqnarray}
 Here $C_A$ and $C_B$  are coefficients which are not shown explicitly for simplicity.
 In the color superconductivity theory,   the one-gluon exchange induced  Nambu--Jona-Lasinio like Models
 will also lead to the $S^a-\bar{S}^a$ type and $P^a-\bar{P}^a$ type diquark
 pairs \cite{Cahill89},
 \begin{equation}
  G \bar{q}\gamma^\mu \frac{\lambda^a}{2} q \bar{q}\gamma_\mu \frac{\lambda^a}{2}
  q \propto C_A S^a \bar{S}^a + C_B P^a \bar{P}^a+\cdots \, .
 \end{equation}   So we can  take the point of view that  the lowest
lying scalar mesons are $S$-wave bound states of
diquark-antidiquark pairs of  $S^a-\bar{S}^a$ type and $P^a-\bar{P}^a$
type.

In this article, we investigate the masses of the scalar nonet
mesons  with two interpolating currents
 respectively and choose the following two-point correlation functions,
\begin{eqnarray}
\Pi_S^i(p)=i\int d^4x ~e^{ip.x}\langle 0
|T[J_S^i(x){J^i_S}^\dagger(0)]|0\rangle.
\end{eqnarray}
Here the  current  $J_S^i$ denotes $J^A_{f_0}$, $J^B_{f_0}$,
$J^A_{a_0^0}$ , $J^B_{a_0^0}$, $J^A_{a^+_0}$ , $J^B_{a^+_0}$,
$J^A_{\kappa^+}$ , $J^B_{\kappa^+}$, $J^A_{\kappa^0}$ ,
$J^B_{\kappa^0}$, $J^A_{\sigma}$ and $J^B_{\sigma}$. According to
the basic assumption of current-hadron duality in the QCD sum
rules approach \cite{Shifman79}, we insert  a complete series of
intermediate states satisfying the unitarity   principle with the
same quantum numbers as the current operator $J_S^i(x)$
 into the correlation functions in
Eq.(16)  to obtain the hadronic representation. Isolating the
ground state contributions from the pole terms of the nonet
mesons, we get the  result,
\begin{eqnarray}
\Pi_S^i(p)=\frac{2f_S^{i2}m_s^{i8}}{m_S^{i2}-p^2}+\cdots \, ,
\end{eqnarray}
where the following definitions have been used,
\begin{equation}
 \langle 0 | J_S^i|S\rangle =\sqrt{2}f_S^im^{i4}_S \;.
 \end{equation}
We have not shown the contributions from the higher resonances and
continuum states explicitly for simplicity.

The  calculation of  operator product expansion in the  deep
Euclidean space-time region is
  straightforward and tedious, technical details are neglected for
  simplicity.  In this article, we consider the vacuum condensates up to dimension six.
  Once  the analytical  results are obtained,
  then we can take the current-hadron dualities below the thresholds
$s_0$ and perform the Borel transformation with respect to the
variable $P^2=-p^2$, finally we obtain  the following sum rules,
\begin{eqnarray}
2f^{A2}_{f_0(a_0)}m^{A8}_{f_0(a_0)}e^{-\frac{m_{f_0(a_0)}^{A2}}{M^2}}=AA , \\
2f^{B2}_{f_0(a_0)}m^{B8}_{f_0(a_0)}e^{-\frac{m_{f_0(a_0)}^{B2}}{M^2}}=BB, \\
2f^{A2}_{\kappa(\kappa^0)}m^{A8}_{\kappa(\kappa^0)}e^{-\frac{m_{\kappa(\kappa^0)}^{A2}}{M^2}}=CC , \\
2f^{B2}_{\kappa(\kappa^0)}m^{B8}_{\kappa(\kappa^0)}e^{-\frac{m_{\kappa(\kappa^0)}^{B2}}{M^2}}=DD, \\
2f^{A2}_{\sigma}m^{A8}_{\sigma}e^{-\frac{m_{\sigma}^{A2}}{M^2}}=EE , \\
2f^{B2}_{\sigma}m^{B8}_{\sigma}e^{-\frac{m_{\sigma}^{B2}}{M^2}}=FF,
\end{eqnarray}
\begin{eqnarray}
AA&=&\int_{4m_s^2}^{s_0}ds e^{-\frac{s}{M^2}}\left\{
\frac{s^4}{2^9 5! \pi^6}+\frac{\langle \bar{s}s\rangle\langle
\bar{q}q\rangle s}{12\pi^2} +\frac{3\langle \bar{q}g_s \sigma G
q\rangle-\langle
\bar{s}g_s \sigma  G s\rangle}{2^6 3 \pi^4}m_s s \right. \nonumber\\
&&\left.-\frac{2\langle \bar{q} q\rangle-\langle \bar{s}
s\rangle}{2^6 3 \pi^4}m_s s^2 +\frac{s^2}{2^9 3 \pi^4} \langle
\frac{\alpha_s GG}{\pi}\rangle \right\}  ,  \\
BB&=&\int_{4m_s^2}^{s_0}ds e^{-\frac{s}{M^2}}\left\{
-\frac{s^4}{2^9 5! \pi^6}+\frac{\langle \bar{s}s\rangle\langle
\bar{q}q\rangle s}{12\pi^2} +\frac{3\langle \bar{q}g_s \sigma  G
q\rangle+\langle
\bar{s}g_s \sigma  G s\rangle}{2^6 3 \pi^4}m_s s \right. \nonumber\\
&&\left.-\frac{2\langle \bar{q} q\rangle+\langle \bar{s}
s\rangle}{2^6 3 \pi^4}m_s s^2 -\frac{s^2}{2^9 3 \pi^4} \langle
\frac{\alpha_s GG}{\pi}\rangle \right\} ,  \\
CC &=&\int_{m_s^2}^{s_0}ds e^{-\frac{s}{M^2}}\left\{
\frac{s^4}{2^9 5! \pi^6}+\frac{\langle \bar{q}q\rangle^2+\langle
\bar{s}s\rangle\langle \bar{q}q\rangle }{24\pi^2}s +\frac{3\langle
\bar{q}g_s \sigma G q\rangle-\langle
\bar{s}g_s \sigma  G s\rangle}{2^7 3 \pi^4}m_s s \right. \nonumber\\
&&\left.-\frac{2\langle \bar{q} q\rangle-\langle \bar{s}
s\rangle}{2^7 3 \pi^4}m_s s^2 +\frac{s^2}{2^9 3 \pi^4} \langle
\frac{\alpha_s GG}{\pi}\rangle \right\}  ,  \\
DD&=&\int_{m_s^2}^{s_0}ds e^{-\frac{s}{M^2}}\left\{-
\frac{s^4}{2^9 5! \pi^6}+\frac{\langle \bar{q}q\rangle^2+\langle
\bar{s}s\rangle\langle \bar{q}q\rangle }{24\pi^2}s +\frac{3\langle
\bar{q}g_s \sigma G q\rangle+\langle
\bar{s}g_s \sigma  G s\rangle}{2^7 3 \pi^4}m_s s \right. \nonumber\\
&&\left.-\frac{2\langle \bar{q} q\rangle+\langle \bar{s}
s\rangle}{2^7 3 \pi^4}m_s s^2 -\frac{s^2}{2^9 3 \pi^4} \langle
\frac{\alpha_s GG}{\pi}\rangle \right\}  ,   \\
EE&=&\int_{0}^{s_0}ds e^{-\frac{s}{M^2}}\left\{ \frac{s^4}{2^9 5!
\pi^6}+\frac{\langle \bar{q}q\rangle^2}{12\pi^2}s +\frac{s^2}{2^9
3 \pi^4} \langle
\frac{\alpha_s GG}{\pi}\rangle \right\}  ,  \\
FF&=&\int_{0}^{s_0}ds e^{-\frac{s}{M^2}}\left\{- \frac{s^4}{2^9 5!
\pi^6}+\frac{\langle \bar{q}q\rangle^2}{12\pi^2}s -\frac{s^2}{2^9
3 \pi^4} \langle \frac{\alpha_s GG}{\pi}\rangle \right\} ,
\end{eqnarray}
here we have taken  the same notation  $s_0$ for the threshold
parameters $s^0_{f_0(a_0)}$ ,
 $s^0_{\kappa^+(\kappa^0)}$ and $s^0_{\sigma}$.  Differentiate the above sum rules with respect to the
variable $\frac{1}{M^2}$, then eliminate the quantities
$f^A_{f_0(a_0)}$ , $f^B_{f_0(a_0)}$ , $f^A_{\kappa^+(\kappa^0)}$ ,
$f^B_{\kappa^+(\kappa^0)}$, $f^A_{\sigma}$ and $f^B_{\sigma}$, we
obtain
\begin{eqnarray}
m^{A2}_{f_0(a_0)}&=&\int_{4m_s^2}^{s_0}ds
e^{-\frac{s}{M^2}}\left\{ \frac{s^5}{2^9 5! \pi^6}+\frac{\langle
\bar{s}s\rangle\langle \bar{q}q\rangle s^2}{12\pi^2}
+\frac{3\langle \bar{q}g_s \sigma  G q\rangle-\langle
\bar{s}g_s \sigma  G s\rangle}{2^6 3 \pi^4}m_s s^2 \right. \nonumber\\
&&\left.-\frac{2\langle \bar{q} q\rangle-\langle \bar{s}
s\rangle}{2^6 3 \pi^4}m_s s^3 +\frac{s^3}{2^9 3 \pi^4} \langle
\frac{\alpha_s GG}{\pi}\rangle \right\}/AA, \\
m^{B2}_{f_0(a_0)}&=&\int_{4m_s^2}^{s_0}ds
e^{-\frac{s}{M^2}}\left\{ -\frac{s^5}{2^9 5! \pi^6}+\frac{\langle
\bar{s}s\rangle\langle \bar{q}q\rangle s^2}{12\pi^2}
+\frac{3\langle \bar{q}g_s \sigma  G q\rangle+\langle
\bar{s}g_s \sigma  G s\rangle}{2^6 3 \pi^4}m_s s^2 \right. \nonumber\\
&&\left.-\frac{2\langle \bar{q} q\rangle+\langle \bar{s}
s\rangle}{2^6 3 \pi^4}m_s s^3 -\frac{s^3}{2^9 3 \pi^4} \langle
\frac{\alpha_s GG}{\pi}\rangle \right\} /BB, \\
m^{A2}_{\kappa^+(\kappa^0)} &=&\int_{m_s^2}^{s_0}ds
e^{-\frac{s}{M^2}}\left\{ \frac{s^5}{2^9 5! \pi^6}+\frac{\langle
\bar{q}q\rangle^2+\langle \bar{s}s\rangle\langle \bar{q}q\rangle
}{24\pi^2}s^2 +\frac{3\langle \bar{q}g_s \sigma G q\rangle-\langle
\bar{s}g_s \sigma  G s\rangle}{2^7 3 \pi^4}m_s s^2 \right. \nonumber\\
&&\left.-\frac{2\langle \bar{q} q\rangle-\langle \bar{s}
s\rangle}{2^7 3 \pi^4}m_s s^3 +\frac{s^3}{2^9 3 \pi^4} \langle
\frac{\alpha_s GG}{\pi}\rangle \right\}  /CC,  \\
m^{B2}_{\kappa^+(\kappa^0)}&=&\int_{m_s^2}^{s_0}ds
e^{-\frac{s}{M^2}}\left\{- \frac{s^5}{2^9 5! \pi^6}+\frac{\langle
\bar{q}q\rangle^2+\langle \bar{s}s\rangle\langle \bar{q}q\rangle
}{24\pi^2}s^2 +\frac{3\langle \bar{q}g_s \sigma G q\rangle+\langle
\bar{s}g_s \sigma  G s\rangle}{2^7 3 \pi^4}m_s s^2 \right. \nonumber\\
&&\left.-\frac{2\langle \bar{q} q\rangle+\langle \bar{s}
s\rangle}{2^7 3 \pi^4}m_s s^3 -\frac{s^3}{2^9 3 \pi^4} \langle
\frac{\alpha_s GG}{\pi}\rangle \right\}/DD  ,   \\
m^{A2}_{\sigma}&=&\int_{0}^{s_0}ds e^{-\frac{s}{M^2}}\left\{
\frac{s^5}{2^9 5! \pi^6}+\frac{\langle
\bar{q}q\rangle^2}{12\pi^2}s^2 +\frac{s^3}{2^9 3 \pi^4} \langle
\frac{\alpha_s GG}{\pi}\rangle \right\}/EE  ,  \\
m^{B2}_{\sigma}&=&\int_{0}^{s_0}ds e^{-\frac{s}{M^2}}\left\{-
\frac{s^5}{2^9 5! \pi^6}+\frac{\langle
\bar{q}q\rangle^2}{12\pi^2}s^2 -\frac{s^3}{2^9 3 \pi^4} \langle
\frac{\alpha_s GG}{\pi}\rangle \right\}/FF  .
\end{eqnarray}
 It is easy to perform the   $s$ integral in
 Eqs.(25-36),  we prefer this form for simplicity.

\section{Numerical Results}
In calculation, the parameters are taken as $\langle \bar{s}s
\rangle=0.8\langle \bar{u}u \rangle$, $\langle \bar{s}g_s\sigma  G s
\rangle=m_0^2\langle \bar{s}s \rangle$, $\langle \bar{q}g_s\sigma
 G q \rangle=m_0^2\langle \bar{q}q \rangle$, $m_0^2=0.8GeV^2$, $\langle \bar{u}u
\rangle=\langle \bar{d}d \rangle=\langle \bar{q}q \rangle=(-219
MeV)^3$, $\langle \frac{\alpha_sGG}{\pi} \rangle=(0.33 GeV)^4$,
 $m_u=m_d=0$ and $m_s=150MeV$.  The main contributions to the sum rules
 come from  the quark condensates terms (i.e. $\langle \bar{q}q \rangle$ and $\langle \bar{s}s \rangle$),
 here we have taken the standard
values and neglected the uncertainties, small variations of those
condensates will not  lead to large  changes about   the numerical
 values.  The threshold parameters are taken as $s^0_{f_0(a_0)}=(1.4-1.6)GeV^2$ ,
 $s^0_{\kappa^+(\kappa^0)}=(1.0-1.2)GeV^2$ and $s^0_{\sigma}=(0.8-1.0)GeV^2$ to avoid possible contaminations
 from the  higher resonances and continuum states. The widths of the $f_0(980)$ and
  $a_0(980)$ mesons are narrow, the threshold parameters
  $s^0_{f_0(a_0)}=(1.4-1.6)GeV^2$ are sufficient to include the
  contributions from those mesons. Although the existence of the $\sigma$ meson is
  confirmed, there are still controversy  about its mass and width,
  here we take the point of view that the $\sigma$ meson is the isosinglet  $S$-wave  $\pi\pi$ resonance near
  $600MeV$ and take the largest $s^0_{\sigma}$  to be the $K\bar{K}$
  threshold. As far as the $\kappa(800)$ mesons are concerned, there are
 still controversy about their existence, here we take them as the
 $S$-wave isospin-$\frac{1}{2}$
  $K\pi$ resonance with the Breit-Wigner mass about $800MeV$ and width
  about $400MeV$, our numerical results support this assumption \cite{ReviewScalar}.
In the region $M^2=(1.2-3.0)GeV^2$, the sum rules for
$m^A_{f_0}=m^A_{a_0}$,  $m^B_{f_0}=m^B_{a_0}$,
$m^A_{\kappa^+}=m^A_{\kappa^0}$, $m^B_{\kappa^+}=m^B_{\kappa^0}$,
$m^A_{\sigma}$ and $m^B_{\sigma}$ are almost independent of  the
Borel parameter $M^2$, the values of masses for those mesons are
shown in Table 1.
  Due to the special quark constituents and Dirac structures of the
  interpolating currents, the $f_0(980)$ and $a_0(980)$ ,
  the $\kappa^+(800)$ and $\kappa^0(800)$ have degenerate masses respectively.
   For the $S^a-\bar{S}^a$ type  interpolating currents $J^A_{f_0}$ ,
   $J^A_{a_0^0}$,   $J^A_{a_0^+}$,
    $J^A_{\kappa^+}$ ,  $J^A_{\kappa^0}$ and
    $J^A_{\sigma}$, the values for
   masses are about $m^A_{f_0}=m^A_{a_0}=(0.96-1.02) GeV$, $m^A_{\kappa^+}=m^A_{\kappa^0}=(0.80-0.88)
   GeV$ and $m^A_{\sigma}=(0.72-0.80) GeV$,
  while for the $P^a-\bar{P}^a$ type interpolating
  currents $J^B_{f_0}$ , $J^B_{a_0^0}$,  $J^B_{a_0^+}$,  $J^B_{\kappa^+}$ , $J^B_{\kappa^0}$ and $J^B_{\sigma}$,
   the values for
   masses are about $m^B_{f_0}=m^B_{a_0}=(0.95-1.01) GeV$, $m^A_{\kappa^+}=m^A_{\kappa^0}=(0.79-0.87)
   GeV$ and $m^A_{\sigma}=(0.71-0.79) GeV$.
   In this article, we take the ideal mixing limit for the two isoscalar
   mesons, the $f_0(980)$ and $\sigma(600)$. We can investigate the mixing  with the following
   substitutions for the interpolating currents,
\begin{eqnarray}
J^A_\sigma \rightarrow cos\theta J^A_\sigma -sin\theta J^A_{f_0},
\,\,\, J^A_{f_0} \rightarrow sin\theta J^A_\sigma +cos\theta J^A_{f_0},\nonumber \\
J^B_\sigma \rightarrow cos\varphi J^B_\sigma -sin\varphi
J^B_{f_0}, \,\,\, J^B_{f_0} \rightarrow sin\varphi J^B_\sigma
+cos\varphi J^B_{f_0},
\end{eqnarray}
 here $\theta$ and $\varphi$ are mixing angles.   From above equations, we can obtain
 lower masses for the $f_0(980)$ meson and higher masses for the $\sigma(600)$ meson with small mixing angles,
 which will not potentially change our numerical results.
 There may be some $q\bar{q}$ components in those nonet scalar mesons, as the  $\bar{q}q$ type interpolating currents
 can also give the correct spin and parity, $J^{P}=0^{+}$. We can
 explore the mixing between the two quark and four quark components  by introducing a free parameter $t$
 with   mass dimension  3, which can vary
 between $0$ and $\infty$, for example,
 \begin{eqnarray}
 J^A_{\sigma}\rightarrow J^A_{\sigma} +t\frac{\bar{u}u+\bar{d}d}{2},
 \nonumber \\
J^A_{\kappa^+}\rightarrow J^A_{\kappa^+}+t \bar{s}u.
 \end{eqnarray}
 The analysis based on QCD sum rules approach indicates that the masses of the ground states
 of  $\bar{q}q$  type interpolating currents are always  larger than $ 1GeV$ or about $1 GeV$ \cite{YangMZ},
   small $\bar{q}q$ components will lead to slightly higher masses for those
  scalar mesons. In the limit $t\rightarrow \infty$, we obtain the
  sum rules for the ground states of $\bar{q}q$ type interpolating
  currents.
  Although the  values for masses $m^A_{f_0}$,
  $m^A_{a_0^0}$, $m^A_{a_0^+}$, $m^A_{\kappa^+}$, $m^A_{\kappa^0}$ and $m^A_{\sigma}$
  lie a little above the corresponding masses $m^B_{f_0}$, $m^B_{a_0^0}$, $m^B_{a_0^+}$, $m^B_{\kappa^+}$, $m^B_{\kappa^0}$
   and $m^B_{\sigma}$,
  we can not get to the conclusion that the scalar nonet mesons
  prefer the $S^a-\bar{S}^a$ type interpolating currents $J^A_{f_0}$ , $J^A_{a_0^0}$, $J^A_{a_0^+}$, $J^A_{\kappa^+}$ ,  $J^A_{\kappa^0}$ and
    $J^A_{\sigma}$
  to the $P^a-\bar{P}^a$ type interpolating currents $J^B_{f_0}$ , $J^B_{a_0^0}$, $J^B_{a_0^+}$, $J^B_{\kappa^+}$ , $J^B_{\kappa^0}$ and $J^B_{\sigma}$.
Precise determination of what type interpolating currents we
should choose calls for original theoretical approaches,
 the contributions from  the direct instantons may do the work. In our recent work,
 we observe that the contributions from the  direct instantons are
neglectable  for the pentaquark state $\Theta^+(1540)$
\cite{Wang05}, however,
 the contributions from the direct instantons can improve the QCD sum rule greatly
 in some channels,  for example,  the
nonperturbative contributions from the direct instantons to the
conventional operator product expansion can significantly improve
the stability of chirally odd  nucleon sum rules
\cite{Dorokhov90,Forkel}. Despite whatever the interpolating
currents may be, we observe that they can both give the correct
mass hierarchy pattern of the scalar nonet mesons below $1GeV$,
there must be some four-quark constituents in those mesons.

 \begin{table}[ht]
         \caption{\label{Tablechi}The values of the scalar nonet mesons }
         \begin{center}
         \begin{tabular}{c||c}
         \hline
              $m_{f_0(a_0)}^A=(0.96-1.02)GeV$   &  $s_0=(1.4-1.6) GeV^2$   \\ \hline
    $m_{f_0(a_0)}^B=(0.95-1.01)GeV$     & $s_0=(1.4-1.6)GeV^2$  \\ \hline
     $m_{\kappa^+(\kappa^0)}^A=(0.80-0.88)GeV$   &  $s_0=(1.0-1.2)GeV^2$\\ \hline
    $m_{\kappa^+(\kappa^0)}^B=(0.79-0.87)GeV$    &  $s_0=(1.0-1.2)GeV^2$   \\ \hline
     $m_{\sigma}^A=(0.72-0.80)GeV$       &   $s_0=(0.8-1.0)GeV^2$ \\ \hline
      $m_{\sigma}^B=(0.71-0.79)GeV$       &  $s_0=(0.8-1.0)GeV^2$ \\ \hline
                        \end{tabular}
         \end{center}
         \end{table}

\section{Conclusions}
In this article, we take the point of view that the $0^{++}$ nonet
mesons  below $1GeV$ are four-quark states
$(qq)_{\bar{3}}(\bar{q}\bar{q})_3$ in the ideal mixing limit, and
devote to determine the values of their masses  in the framework
of  the QCD sum rules approach.  Due to the special quark
constituents and Dirac structures of the
  interpolating currents, the  $f_0(980)$ and $a_0(980)$ ,
  the $\kappa^+(800)$ and $\kappa^0(800)$ have degenerate masses respectively.
     For the $S^a-\bar{S}^a$ type interpolating currents $J^A_{f_0}$ ,
      $J^A_{a_0^0}$,  $J^A_{a_0^+}$, $J^A_{\kappa^+}$ ,  $J^A_{\kappa^0}$ and
    $J^A_{\sigma}$, the values for
   masses are about $m^A_{f_0}=m^A_{a_0}=(0.96-1.02) GeV$, $m^A_{\kappa^+}=m^A_{\kappa^0}=(0.80-0.88)
   GeV$ and $m^A_{\sigma}=(0.72-0.80) GeV$,
  while for the $P^a-\bar{P}^a$ type interpolating currents
  $J^B_{f_0}$ , $J^B_{a_0^0}$, $J^B_{a_0^+}$, $J^B_{\kappa^+}$ , $J^B_{\kappa^0}$ and $J^B_{\sigma}$,
   the values for
   masses are about $m^B_{f_0}=m^B_{a_0}=(0.95-1.01) GeV$, $m^A_{\kappa^+}=m^A_{\kappa^0}=(0.79-0.87)
   GeV$ and $m^A_{\sigma}=(0.71-0.79) GeV$.
  Although the  values for masses $m^A_{f_0}$, $m^A_{a_0^0}$, $m^A_{a_0^+}$, $m^A_{\kappa^+}$, $m^A_{\kappa^0}$ and $m^A_{\sigma}$
  lie a little above the corresponding masses $m^B_{f_0}$, $m^B_{a_0^0}$, $m^B_{a_0^+}$, $m^B_{\kappa^+}$, $m^B_{\kappa^0}$
   and $m^B_{\sigma}$,
  we can not get to the conclusion that the scalar nonet mesons
  prefer the $S^a-\bar{S}^a$ type interpolating currents $J^A_{f_0}$ , $J^A_{a_0^0}$, $J^A_{a_0^+}$, $J^A_{\kappa^+}$ ,  $J^A_{\kappa^0}$ and
    $J^A_{\sigma}$
  to the $P^a-\bar{P}^a$ type interpolating currents $J^B_{f_0}$ , $J^B_{a_0^0}$, $J^B_{a_0^+}$, $J^B_{\kappa^+}$ , $J^B_{\kappa^0}$ and $J^B_{\sigma}$.
 Despite whatever the interpolating
currents may be, we observe that  they can both give the correct
mass hierarchy pattern of the scalar nonet, there must be some
four-quark constituents in those mesons, our results support the
four-quark model  and the  hybrid model.  In the  hybrid model,
those mesons are
 four-quark states $(qq)_{\bar{3}}(\bar{q}\bar{q})_3$ in $S$-wave near the
center, with some constituent $q \bar{ q}$ in $P$-wave, but
further out they rearrange into  $(q \bar{ q})_1(q \bar{ q})_1$
states and finally as meson-meson states \cite{Close2002}. Precise
determination of what type interpolating currents we should choose
calls for original theoretical approaches,
 the contributions from  the direct instantons may do the work.

\section*{Acknowledgment}
This  work is supported by National Natural Science Foundation,
Grant Number 10405009,  and Key Program Foundation of NCEPU. The
authors are indebted to Dr. J.He (IHEP) , Dr. X.B.Huang (PKU) and Dr. L.Li (GSCAS)
for numerous help, without them, the work would not be finished. The author would
also thanks Prof. M.Nielsen for helpful discussion.

\end{document}